It has been shown that $t^0_0$ component of the energy-momentum pseudotensor in the case of cylindrically symmetrical static ggravitational field cannot be interpreted as energy density of the gravitation field. An approach has been suggested allows one to express the energy density of the cylindrically or spherically symmetrical static gravitation field in terms of the metric tensor components. The approach based on the consideration of the process of isothermal compression of a cylinder consisted of incoherent matter

# The gravitation energy for a cylindrically and spherically symmetrical system

**Roald Sosnovskiy**
Technical University, 194021, St. Petersburg, Russia
E-mail:rosov2@yandex.ru

It has been shown that $t^0_0$ component of the energy-momentum pseudotensor in the case of cylindrically symmetrical static g gravitational field cannot be interpreted as energy density of the gravitation field. An approach has been suggested allows one to express the energy density of the cylindrically or spherically symmetrical static gravitation field in terms of the metric tensor components. The approach based on the consideration of the process of isothermal compression of a cylinder consisted of incoherent matter.

## 1. Introduction

The relativistic theory of gravitation has an intrinsic contradiction, which arise from the fact that the energy-momentum tensor is a noncovariant value. This contradiction produces some difficulties in the consequent approach to the energy conservation principle. This issue is a matter of discussion up to now [1], [2].

As it is shown below in this paper, the $t^0_0$ component of the energy-momentum pseudotensor for a cylindrically symmetrical gravitation field cannot be interpreted as the energy density of the gravitation field. In this paper the approach is proposed allows one to express the energy density of such a field through the components of a metric tensor. This approach based on the consideration of the isothermal compression of the cylinder consisted of the incoherent matter

It has been shown [1] that the length element for the cylindrically symmetrical system can be represented as

$$dS^2 = g_{00} dT^2 + g_{11} dR^2 + g_{22} d\Phi^2 + g_{33} dZ^2 \qquad (1)$$

For the static field

$$|g_{ii}| = (R/R_0)^{a_i}, \; i = 0,1,3 \; ; \; g_{22} = -R^2 (R/R_0)^{a_2} \qquad (2)$$

Here

$$x^0 = T; \; x^1 = R; \; x^2 = \Phi; \; x^3 = Z; \; a_0 = \frac{4(1-A)}{A^2+3}; \; a_1 = 0$$

$$a_2 = -\frac{2(A-1)^2}{A^2+3}; \; a_3 = \frac{2(A^2-1)}{A^2+3} \qquad (3)$$

and $R_0$ is the radius of the cylinder. Formulas (2), (3) correspond to the zero scalar curvature $R_g=0$ in the whole exterior of the gravitating cylinder.

The choice of the Lagrangian of the external gravitational fields is a matter of some difficulties. It is obvious that the metrical Lagrangian $L \sim \sqrt{g}\, R_g$ is equal to zero in the exterior of the cylinder. It is also the case for $L \sim g^{\mu\nu}(\Gamma^{\alpha}_{\mu\nu}\Gamma^{\beta}_{\alpha\beta} - \Gamma^{\alpha}_{\beta\mu}\Gamma^{\beta}_{\alpha\nu})$ and the formal energy-momentum pseudotensor density tensor [4]

$$t^{\nu}_{\mu} = g_{\alpha\beta,\mu}\, \partial L/\partial g_{\alpha\beta,\nu} - g^{\nu}_{\mu} L$$

is equal to zero.

On the other hand, the energy of the test particles moving in the external gravitational field of cylinder is not constant. As well as the only source of the energy variation of the test particle is the gravitation field energy, the value $t^{0}_{0}$ cannot be considered as a density w of this energy. Moreover, the value $t^{0}_{0}$ does not fulfill the correspondence principle (relativistic) [5], according to which in the limiting case of weak field all equations should grade into equations of the Newton theory. In particular the energy density of the field of the cylinder should transform into $w_N = -GM_z^2/2\pi r^2$ ($M_z$ is a linear mass density of a cylinder).

Another form of the $t^{\nu}_{\mu}$ has been suggested in [6]:

$$-\frac{16\pi G}{c^4} g t^{ik} = b^{ik}_{,l} b^{lm}_{,m} - b^{il}_{,l} b^{km}_{,m} + \tfrac{1}{2} g^{ik} g_{lm} b^{ln}_{,p} b^{pm}_{,n} - \left(g^{il} g_{mn} b^{kn}_{,p} b^{mp}_{,l} + g^{kl} g_{mn} b^{in}_{,p} b^{mp}_{,l}\right) + \\
+ g_{lm} g^{np} b^{il}_{,n} b^{km}_{,p} + \tfrac{1}{8}\left(2 g^{il} g^{km} - g^{ik} g^{lm}\right)\left(2 g_{np} g_{qr} - g_{pq} g_{nr}\right) b^{nr}_{,l} b^{pq}_{,m} \tag{4}$$

Here $b^{ik} = \sqrt{-g}\, g^{ik}$ and index ",i" means the "simple" $x^i$ derivative.

It can by easily obtained from (2) and (4) that for a cylinder

$$R^2 t^{00} = -\frac{c^4}{16\pi G}\left(\frac{R}{R_0}\right)^{-a_0}\left(2 - 3a_0 + a_0^2\right) \tag{5}$$

It can be seen that in this case $t^{0}_{0} = g_{00} t^{00} \neq 0$, but in the limiting case $M_z \to 0$ (no graviting mass at all) $g_{00} \to 1$ and $a_0 \to 0$. The density of field energy in this limiting case should be zero, but according to (5) it does not. Therefore, $t^{0}_{0}$ from (4) also cannot be interpreted as a "true" density of gravitational field energy.

Nevertheless the expression for the density of the gravitational field energy is necessary for the consideration of the set of problems, for example in the global strings problem [7]. In this paper some approach for obtaining of such expression is proposed. This approach can also be applied to spherically symmetrical system.

### 2. The energy density of the static gravitation field.

The approach proposed in this paper based on the consideration of process of the isothermal stagewise compression of the hollow cylinder consisted of incoherent matter which produces gravitational field. The decreasing of the cylinder radius is considered to be result of the consecutive displacement of the symmetrical infinitesimal layers formed by particles. The self-field dependence of the layer movement can be neglected. The process considered to be isothermal, what means that the energy related to the particles movement dissipated.

The expressions we are going to derive should fulfill following requirement:

·(a) in the absence of mass energy density should be equal to zero

·(b) the correspondence principle should be fulfilled including the energy part

·(c) the local energy conservation law should be fulfilled.

The (c) requirement in the limiting case of small masses is a corollary of (b) and we suppose that it is correct for large mass if the field energy is localized.

Let us consider the displacement of the particles layer from position $x^1 = x^1_1$ to the position $x^1 = x^1_1 + dx^1$, $dx^1 < 0$. The free particles motion equations are

$$\frac{d}{d\tau}\left(\frac{\partial L}{\partial \dot{x}^\mu}\right) - \frac{\partial L}{\partial x^\mu} = 0 \tag{6}$$

where $\tau$ is the intrinsic time, $\dot{x}^\mu = \frac{dx^\mu}{d\tau}$ and

$$L(\dot{x}^\sigma, x^\sigma) = \frac{1}{2} g_{\mu\nu} \dot{x}^\mu \dot{x}^\nu \tag{7}$$

For the radial movement $\dot{x}^0 = \dot{T}^0 \neq 0$, $\dot{x}^1 = \dot{R} \neq 0$, $\dot{x}^2 = \dot{x}^3 = 0$. It can be shown from (1) and (7) that

$$c^2 = g_{00} \dot{x}^{02} + g_{11} \dot{x}^{12} \tag{8}$$

$$L = \frac{1}{2} g_{00} \dot{x}^{02} + \frac{1}{2} g_{11} \dot{x}^{12} \tag{9}$$

where $g_{ii} = g_{ii}(x^1)$ due to (2). For $\mu = 0$ and initial conditions $x^1 = x^1_1$ and $dx^1/dx^0 = 0$ (6) - (9) lead to

$$\frac{dx^1}{dx^0} = -\sqrt{\left[\frac{g_{00}(x^1)}{g_{00}(x^1_1)} - 1\right]\frac{g_{00}(x^1)}{g_{11}(x^1)}} \tag{10}$$

The energy of the particle with rest mass $\delta m$ can be obtained from the relation [1]

$$E = \delta m\, c^2\, u^i\, \frac{g_{0i}}{\sqrt{g_{00}}}, \quad u^i = \frac{dx^i}{c d\tau} \tag{11}$$

For $i = 0$ in (11), the change in energy near $x^1_1$ can be obtained from formulas (10) and (11) as

$$dE = -\frac{\delta m c^2 g_{00,1}(x^1_1)}{2 g_{00}(x^1_1)} dx^1 \tag{12}$$

The whole mass of the gravitating matter can be expressed in field produced by it as [1]:

$$M = \frac{c^2}{4\pi G} \oint \sqrt{-g}\, g^{0n} \Gamma^\alpha_{0n} dS_\alpha \tag{13}$$

For the cylindrical symmetry $\alpha = 1$, $n = 0$, $\Gamma^1_{00} = -\frac{1}{2} g^{11} g_{00,1}$ and using components (2) one can be obtain from (13) for the cylindrical surface

$$a_0 = 4GMz/c^2 \tag{14}$$

This parameter does not depend on the mass distribution over R. Equations (8), (9), (11) and (14) lead to the equation for the energy of the free-moving particle which satisfy the (b) requirement

$$\frac{E-\delta mc^2}{\delta m}=c^2\left[\sqrt{\frac{g_{00}(R_1,M_z)}{g_{00}(R,M_z)}}-1\right]\approx 2GM_z\ln\frac{R_1}{R}=\Delta U \qquad (15)$$

Here U is the Newton potential.

Let mass $M_z$ consider to be uniformly distributed over the cylindrical surface with radius $R_2$. In the area $R < R_2$ there is no field. The layers discussed above move one by one from this surface. The linear mass density consider to be $dM_z$. When the layer passes the area dR field changes only in cylindrical area between R and R-dR because parameter $a_0$ does not depend on $R_0$, but it depends on the linear mass density inside the current area with radius R. The requirement (c) results in $dE + dE_f = 0$ where $dE_f$ is the change of field energy over the unit length of the cylinder. When $\delta m = dM_{z1}$ it can be seen from (12) and (15) that

$$d^2 E_f = -\frac{dM_{z1}c^2}{2}\frac{g_{00,1}(R,M_{z1})}{g_{00}(R,M_{z1})}dR \qquad (16)$$

where $M_{z1}$ is the current linear density of mass inside the area with radius R. The energy $dE_f$ is localized in the layer dR. Sequentially moving mass $M_z$ from $R_2$ to $R_0$ in this way we can get the field energy in the cylindrical layer $R_0 \le R \le R_2$ as

$$E_f = -\frac{c^2}{2}\int_{R_0}^{R_2}dR\int_0^{M_z}\frac{g_{00,1}(R,M_{z1})}{g_{00}(R,M_{z1})}dM_{z1} = -\frac{c^2}{2}\int_0^{M_z}\ln\frac{g_{00}(R_2,M_{z1})}{g_{00}(R_0,M_{z1})}dM_{z1} \qquad (17)$$

or taking into account (2) and (14),

$$M_{fz}=\frac{E_f}{c^2}=-\frac{GM_z^2}{c^2}\ln(R_2/R_0)$$

From (16) the field energy density $w(M_z,R)$ is

$$w(M_z,R)=-\frac{c^2}{2S_1}\int_0^{M_z}\frac{g_{00,1}(R,M_{z1})}{g_{00}(R,M_{z1})}dM_{z1} \qquad (18)$$

where $S_1$ is the lateral area of the surface R = Const per length unit [12]

$$S_1=\int\sqrt{g}\,dx^2 dx^3 \qquad (19)$$

Formula (18) can be rewritten using (2), (14) and (19) as

$$w(M_z,R)=-\frac{GM_z^2}{2\pi R^2} \qquad (20)$$

Thus $w(M_z,R)$ is equal to the Newton's field energy density. This result can be connected with fact, that scalar curvature $R_g=0$ and the space-time is flat. Obviously, the requirement (b) is fulfilled. It should be mention that (1) does not restrict the generality of the result obtained because any initial coordinate system in the system with cylindrical symmetry can be transformed so that (1) is fulfilled.

There are good reasons to take into account the intrinsic energy of gravitational field. Particularly, in [8] the gravitating mass of the central body for the spherically symmetrical system has been expressed as the sum of the own mass of the body and its "gravitational excess", i.e. the mass of the field. Nevertheless in the exterior of the body the energy of the field has not been considered as a sours of the field.

The consistent implementation of the approach to the whole energy of both the matter and the field energy as a source of field has been provided [9] for the Newton gravitation. In [10] the energy-momentum tensor has been considered as the source of metric. The approach

discussed above in the paper can also be applied for such consideration. In this case (16) become

$$M_{fz}(R_2, M_z) = -\frac{1}{2}\int_0^{M_z} \ln\frac{g_{00}(R_2, M_{z1} + M_{fz})}{g_{00}(R_0, M_{z1})} dM_{z1} \qquad (21)$$

where $M_{fz}(R_2, M_z) = E_f(R_2, M_z)/c^2$ is the field mass in the area $R_0 < R < R_2$ and $M_z$ is the rest mass of the unit length of the cylinder. The equation (21) results in

$$M_{fz}(R_2, M_z) = \frac{1-e^{-\nu M_z}}{\nu} - M_z; \quad \nu = (2G/c^2)\ln(R_2/R_0)$$

When $R_2 \to \infty$ $M_{fz} \to -M_z$ so that the field energy is finite and the whole energy of both matter and field tend to 0. However differences between $M_{fz} + M_z$ and $M_z$ within the bounds of the cosmological horizon $R < 10^{28}$ cm become noticeable only when $M_z > 10^{24}$ g/cm. By such great distances the dependence of gravity from distance may be changed [11].Therefore the conduct of quantities by $R_2 \to \infty$ it is necessary to consider with the prudence

### 3. Spherical symmetry

For the spherical symmetry (2) substituted by

$$g_{00} = 1-2GM/c^2R; \; g_{11} = -(1-2GM/c^2R)^{-1}; \; g_{22} = -R^2; \; g_{33} = -R^2\sin^2\theta \qquad (22)$$

where M is the whole mass of spherically distributed matter inside the sphere with radius R. The gravitation field in the exterior of the gravitating sphere does not depends of the R-distribution of the mass inside the sphere. In this maining a spherical field is similar to the cylindrical one. Thas, the equations (8), (9), (12), (15), (17), (18) can be applied to a spherically symmetrical field if one substitute the masses of the unit length of cylinder $M_z$, $M_{fz}$ by the mass of the interior of a sphere M, $M_f$. Then (15), (22) gives the following expression for the energy of a free particle

$$\frac{\Delta E}{\delta m} = c^2\left(\sqrt{\frac{1-2GM/c^2R_1}{1-2GM/c^2R}} - 1\right) \approx GM\left(\frac{1}{R} - \frac{1}{R_1}\right) = \Delta U \qquad (23)$$

where U is the Newton potential.

From (18) and (22) the energy density of the spherically symmetrical field can be represented as

$$w = \frac{c^4}{16\pi GR^2}\left[\ln\left(1 - \frac{2GM}{c^2R}\right) + \frac{2GM}{c^2R}\right] \approx -\frac{GM^2}{8\pi R^4} \qquad (24)$$

For the whole exterior of the body ($R_2 \to \infty$) mass of the field is equal to

$$M_f(\infty, M) = -\frac{M}{2}\left[1 - (1 - c^2R_0/2GM)\ln(1 - 2GM/c^2R_0)\right] \approx -GM^2/2c^2R_0.$$

All formulas above fulfill the correspondence principle (b).

The assumption that the whole energy of field and matter as the source of field and (21),(23) lead to

$$M_f(R_2, M) = -\frac{1}{2}\int_0^M dM' \ln\frac{1 - 2G[M' + M_f(R_2, M')]/c^2 R_2}{1 - 2GM'/c^2 R_0}$$

The field mass is essential comparing with the central body mass when $R_0$ is about the gravitational radius.

## Références